\documentclass[aps,pra,superscriptaddress,twocolumn,longbibliography]{revtex4-1}
\usepackage{mathrsfs}
\usepackage{amsfonts}
\usepackage{amssymb}
\usepackage{amsmath}
\usepackage{graphicx}
\usepackage{lipsum} 
\usepackage{ulem}
\usepackage{sidecap}
\usepackage{color}
\usepackage{tikz}
\usepackage[colorlinks=true, urlcolor=blue, citecolor=blue,linkcolor=blue,citebordercolor={1 0 0},linkbordercolor={0 0 1}]{hyperref}
\usepackage{subeqnarray}
\usepackage{verbatim}
\usepackage{euscript} 

\usepackage[columnwise]{lineno}

\usepackage{array}
\usepackage{multirow}

\renewcommand{\eqref}[1]{Eq.~(\ref{#1})} 

\definecolor{mydarkblue}{rgb}{0,0.08,0.45}
\definecolor{myfavblue}{rgb}{0.1176, 0.392, 1.0}
\hypersetup{
    colorlinks=true,
    linkcolor=mydarkblue,
    citecolor=mydarkblue,
    filecolor=mydarkblue,
    urlcolor=mydarkblue}

\newcommand{\utchem}{Department  of  Chemistry,  University  of  Toronto,  Toronto,  Ontario  M5G 1Z8,  Canada}
\newcommand{\utcomp}{Department  of  Computer Science,  University  of  Toronto,  Toronto,  Ontario  M5S 2E4,  Canada}
\newcommand{\vectorinst}{Vector  Institute  for  Artificial  Intelligence,  Toronto,  Ontario  M5S  1M1,  Canada}
\newcommand{\cifar}{Canadian  Institute  for  Advanced  Research,  Toronto,  Ontario  M5G  1Z8,  Canada}
\newcommand{\accel}{Acceleration Consortium,  Toronto,  Ontario  M5S  3H6,  Canada}

\begin{document}

\title{Atom-by-atom protein generation and beyond with language models}

\author{Daniel Flam-Shepherd}
\affiliation{\utcomp}
\affiliation{\vectorinst}

\author{Kevin Zhu}
\affiliation{\utcomp}
\affiliation{\vectorinst}

\author{Al\'an Aspuru-Guzik}
\affiliation{\utcomp}
\affiliation{\vectorinst}
\affiliation{\utchem}
\affiliation{\cifar}
\affiliation{\accel}

\begin{abstract}
Protein language models learn powerful representations directly from sequences of amino acids. However, they are constrained to generate proteins with only the set of amino acids represented in their vocabulary. In contrast, chemical language models learn atom-level representations of smaller molecules that include every atom, bond, and ring.
In this work, we show that chemical language models can learn atom-level representations of proteins enabling protein generation unconstrained to the standard genetic code and far beyond it. 
In doing so, we show that language models can generate entire proteins atom by atom-- effectively learning the multiple hierarchical layers of molecular information that define proteins from their primary sequence to their secondary, and tertiary structure. 
We demonstrate language models are able to explore beyond protein space-- generating proteins with modified sidechains that form unnatural amino acids. Even further, we find that language models can explore chemical space and protein space simultaneously and generate novel examples of protein-drug conjugates. The results demonstrate the potential for biomolecular design at the atom level using language models. 
\end{abstract}

\maketitle


Proteins are essential components of all life on Earth and are involved in every cellular process. As a result, protein engineering is one of the most important areas of scientific discovery. Significant progress has been made and proteins have been engineered for therapies against viruses \cite{diskin2011increasing} and cancer \cite{adams2005monoclonal}, as well as to alleviate genetic diseases directly \cite{anzalone2019search, gaudelli2017programmable, komor2016programmable}. Artificial intelligence has enormous potential to accelerate scientific progress and automate protein engineering. Already it has led to a breakthrough in highly accurate protein structure prediction \cite{jumper2021highly}. In particular, language models have already begun to have a major impact on protein design  \cite{alley2019unified, hie2023efficient, madani2023large}. 

The important functions proteins carry out and the structure responsible for them originate in the patterns of amino acids in their primary sequence.
Indeed, most language models represent proteins using sequences of amino acids \cite{alley2019unified, hie2023efficient, madani2023large}. However, this ignores atom-level interactions, precluding the model from representing any atom-level protein modification. Allowing for atom-level representations would enable protein generation outside of the genetic code and allow language models to explore an expanded space of biomolecules. Specifically, this would make it possible for the model to propose new unnatural side chains, attach small molecules, and generate linkers between residues that form large macrocycles. 

To learn atom-level representations we must turn to chemical language models, however, these models are typically used for smaller drug-like molecules. Similar to their protein variants-- chemical language models are deep neural networks trained using masking or next-token prediction \cite{flam2022language} but use atom-level linear sequences parsed from molecular graphs \cite{gomez2018automatic, flam2022language}. 
These sequences completely represent the molecule including all atoms, bonds, rings, aromaticity, branching, and stereochemistry. The two most prominent sequence representations are SMILES strings \cite{weininger1988smiles} or SELFIES strings \cite{krenn2019selfies} which are completely robust and always valid. 

Recently, chemical language models \cite{flam2022language} were found to have the ability to generate larger, complex molecules, relative to small drug-like molecules such as the largest molecules in PubChem. These molecules are still much smaller than proteins, but this indicates that atom-level protein generation with language models is feasible. In this work, we demonstrate that chemical language models are capable of generating entire proteins atom by atom, including biomolecules beyond protein space. Specifically, we train models on various biomolecules including proteins from the protein databank. We also create two other synthetic biomolecular datasets, first modified proteins with unnatural amino acids, and proteins with small molecule attachments-- specifically single domain antibodies (sdAbs) from the antibody structural database \cite{constantinnano} attached to molecules from the ZINC dataset \cite{irwin2005zinc}. '

\begin{figure*}[t]
    \centering
    \includegraphics[width=1.0\textwidth]{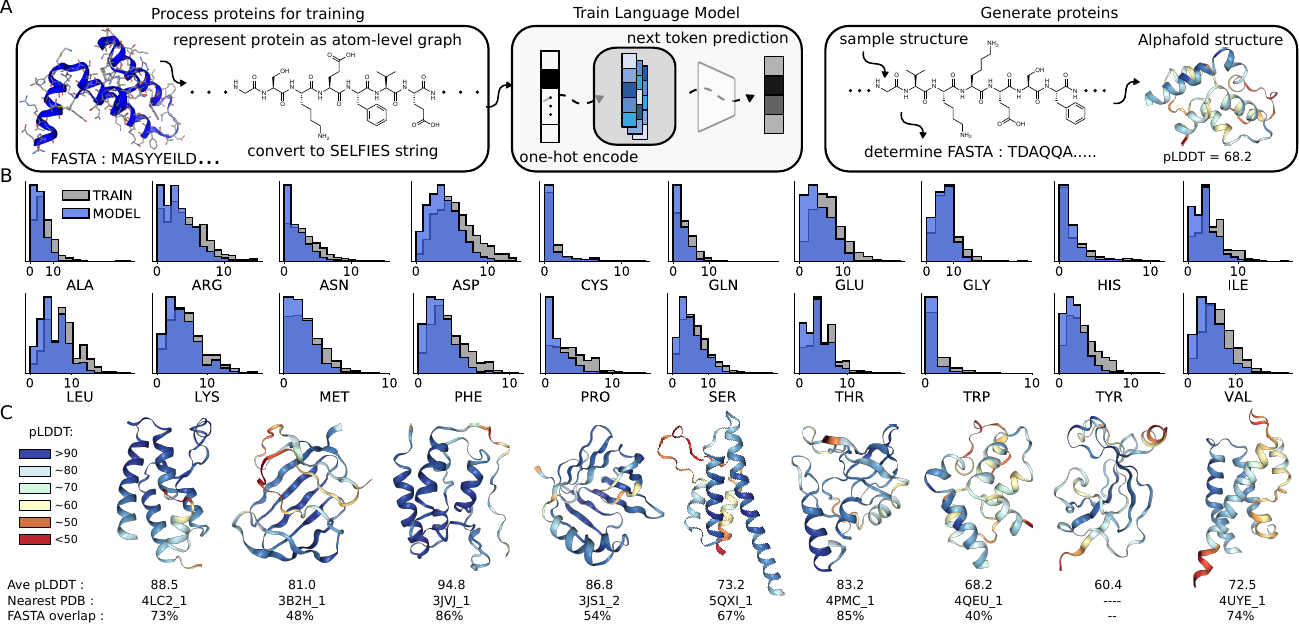}
    \caption{\textbf{Proteins} 
    (A) Dataset preparation. The training workflow for the model: training, generation, amino acid sequence determination, and AlphaFold visualization. 
    (B) Comparison of amino acid distributions. 
    (C) AlphaFold visualizations of model-generated proteins coloured by pLDDT, including the PDB ID of the closest protein and its \% sequence overlap. }
    \label{fig:figure1}
\end{figure*}

We discover that chemical language models can learn the language of proteins entirely from scratch-- by learning to generate atom-level sequences that define proteins with valid primary sequences that correspond to meaningful secondary, and tertiary structure, which we check using AlphaFold \cite{jumper2021highly} structure predictions. Importantly, the language model learns valid protein backbones and natural amino acid structures as well as the primary sequence patterns in the training proteins.  We further demonstrate that language models can generate beyond the standard genetic code-- proteins with novel sidechains that are more complex than the set of standard amino acids. Additionally, we also show that chemical language models can generate novel proteins and small molecules together at the same time as protein-drug conjugates. In particular, we find that the model learns both the protein space of the single domain antibodies and the chemical space defined by the ZINC molecules-- generating antibody-drug conjugates with valid and novel protein sequences and structures attached to novel drug-like molecules warheads similar to the structures in ZINC.   

\section*{Results}

\noindent

In this study, the datasets are constructed by using small proteins from the Protein Data Bank (PDB), specifically between 50 and 150 residues. We use atom-level graph representations of each protein so that sidechain modifications can be made directly. For training, each protein can be parsed to a linear string representation, and random data augmentation can be used to increase the training data size. We describe the main details and results for each dataset in the following sections. 

\begin{figure*}[t]
    \centering
    \includegraphics[width=0.99\textwidth]{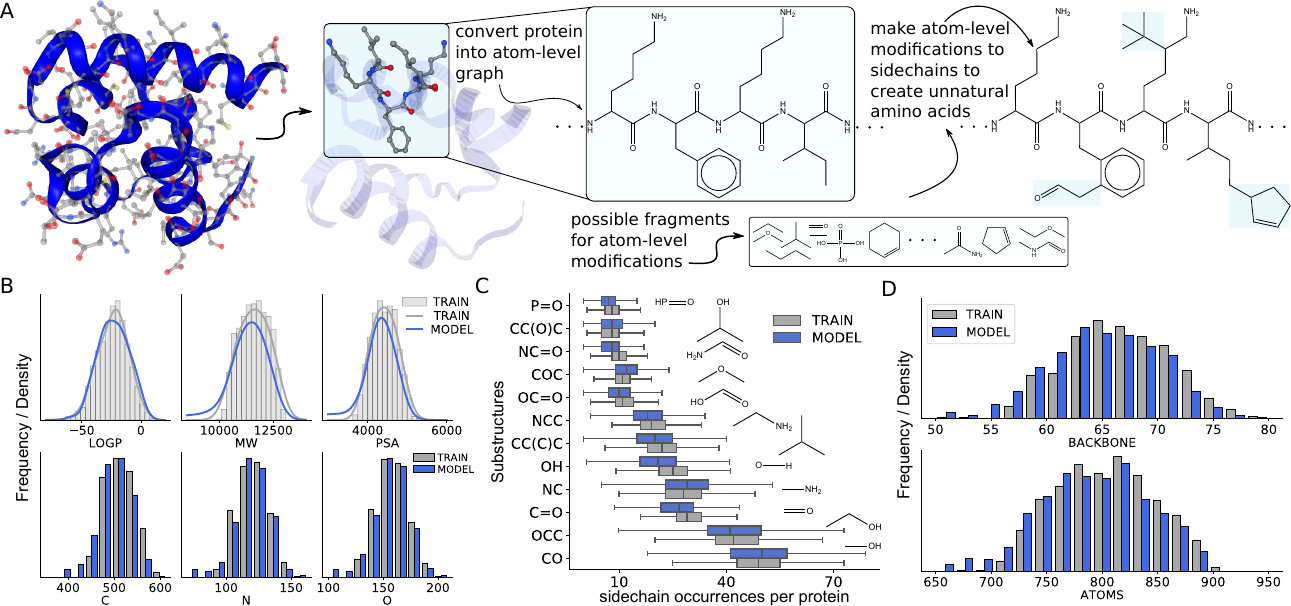}
    \caption{\textbf{Proteins with unnatural amino acids}  
    (A) Dataset of unnatural proteins built by random side chain modification.
    (B) Histograms and density plots of atom-level measures or properties. Density plots use a Gaussian kernel density estimator (KDE) that is fit to  LogP, MW, and PSA values of the training molecules by tuning the bandwidth parameter.
    (C) Boxplots measuring the number of occurrences of small fragments in sidechains of model and training proteins.
    (D) Histograms of backbone size and heavy atom number per protein for model and training proteins.
    }
    \label{fig:figure2}
\end{figure*}

\subsection{Proteins}

For the first dataset, which consists of proteins with standard amino acids and no sidechain modifications, we test the ability of the language model to explore protein space while maintaining protein structure and constraints. After training, as shown in Figure \ref{fig:figure1}(A))-- we generate a thousand (1K) samples from the language model and evaluate their atom, residue, and protein-level properties. At the protein level, we determine if generated samples are proteins and attempt to determine their primary sequence. If we can ascertain their primary sequence, we can use AlphaFold2 \cite{jumper2021highly} to further evaluate if the model has learned the amino acid sequences that correspond to good structure predictions. Additionally, we study model samples for their distribution of amino acids and other atom-level properties that can be computed using rdkit \cite{landrum2013rdkit}.

We check if molecules generated by the model are actually proteins by analyzing if they preserve the basic structure of the protein backbone and natural amino acids form. First, we perform a backbone structure search and then attempt to arrange the backbone from the N terminus to the C terminus while simultaneously classifying each sidechain using another substructure search for the standard set of amino acids. If this is successful and there are no discontinuities in the backbone or other side chain errors, then we classify the sample as a protein and parse the amino acid sequence. By this process, we determine roughly $\sim$ 68.2\% of samples are proteins, furthermore, all the parsed amino acid sequences are unique (there are no duplicates and the model isn't repeating specific proteins) and novel (they are different from the training sequences). 

We compare the distribution of amino acids in the training sequences to the distribution learned by the model based on the generated samples. We plot histograms, in Fig. \ref{fig:figure1}(B), displaying the frequency of occurrence of every amino acid in samples from both the model and the training data-- both distributions are very similar and mostly overlap but for some amino acids, the language model slightly underestimates the training frequencies.

Using Alphafold \cite{jumper2021highly}, in Fig. \ref{fig:figure1}(B), we visualize selected examples of proteins generated by the language model. In each sample, residues are color-coded according to pLDDT, which is a per-residue estimate of the model's confidence on a scale from 0 to 100. Regions with pLDDT $>$ 90 are dark blue and have high accuracy. Regions with pLDDT from 90 down to 70 are still expected to be good predictions and are colored light blue that transitions to green (with decreasing confidence). Regions with pLDDT between 50 and 70 are lower confidence and are colored yellow to green. The regions with pLDDT $<$ 50 are not confident and likely disordered-- these are colored red.

On this scale, in Fig. \ref{fig:figure1}(C), we see that the proteins that are generated by the model result in good structure predictions-- ranging between 70 and 90 pLDDT. This indicates that the model can generate proteins with well-defined structures that are not disordered. For a simple baseline comparison, we considered sequences of random amino acids, the structure predictions for these consistently result in disordered proteins with low pLDDT $<50$.

Additionally, in Fig. \ref{fig:figure1}(C), the proteins generated by the language model contain a variety of secondary structures including alpha helices, beta sheets, and omega loops. Globally, the generated proteins combine many of these secondary structures into various and unique domains. We can conclude, based on these samples, and further examples in Supplementary Fig. \ref{fig:sfigure1}, that language models can generate proteins, atom by atom, not just with valid primary sequences but proteins with meaningful secondary and tertiary structure.

Furthermore, the generated proteins are similar to their nearest training examples in the PDB, to show this in Fig. \ref{fig:figure1}(C) and Supplementary Fig. \ref{fig:sfigure1}, under each protein we label the primary sequence percentage overlap between the generated proteins and their most similar PDB training example-- which ranges from 86\% to 40\% with one other generated protein that has no nearest PDB training example. Based on this, it is evident that the model draws heavily from the amino acid sequence patterns in its training data but does not memorize them. 


Also, in Supplementary Fig. \ref{fig:sfigure2}, we plot histograms comparing atom-level properties of the samples generated from the model with the training data. The model roughly approximates the training distribution of atoms but slightly underestimates some properties.

\begin{figure*}[t]
    \centering
    \includegraphics[width=0.99\textwidth]{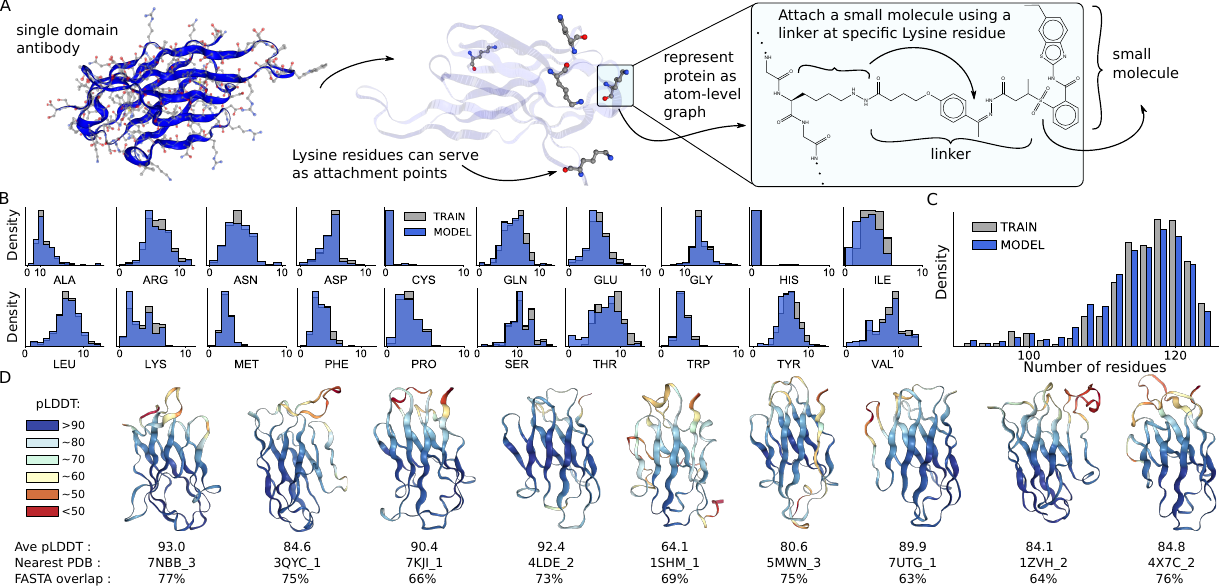}
    \caption{\textbf{Antibody Drug Conjugates -- sdAbs} 
    (A) Single domain Antibody-drug conjugates dataset creation overview.
    (B) Comparison of amino acid distributions for training and model sdAbs.
    (C) Histogram comparing the size of training and model sdAbs (by number of residues).
    (D) Example sdAbs (with warheads excluded) generated by the language model visualized by Alphafold and coloured by pLDDT. 
        Under each, we include the PDB ID of the closest protein and its \% sequence overlap. 
    }
    \label{fig:figure3}
\end{figure*}

\subsection{Proteins with unnatural amino acids}

The next dataset, whose construction is depicted in Fig \ref{fig:figure2}(A), consists of proteins that have random sidechain modifications creating proteins with unnatural amino acids. For this dataset, we select a subset of smaller proteins with 50 to 80 residues from the previous protein dataset. We then modify the selected protein by attaching a randomly chosen small fragment using a random attachment point on every sidechain. This produces a dataset of proteins that are entirely comprised of "unnatural" amino acids. We train language models on these unnatural proteins in order to test the ability of chemical language models to generate biomolecules beyond protein space. Further details about building the training data can be found in the Methods section. Additionally, an example of the unnatural amino acids from a single training protein and model protein can be found in Supplementary Fig. \ref{fig:sfigure3}.  

After training, we again generate 1K samples from the language model for evaluation-- the results are shown in Fig \ref{fig:figure2}, where we test the model's ability to capture atom-level and sidechain-level properties of the unnatural proteins. First, in Fig \ref{fig:figure2} (B), we see that the model learns the continuous atom-level properties of the training proteins including octanol-water partition coefficient (LogP) \cite{wildman1999prediction}, exact molecular weight (MW) and the topological polar surface area (PSA), in addition to learning the number of carbon, nitrogen, and oxygen. Then in Fig \ref{fig:figure2} (C), we see the model learns a similar sidechain structure to the training sidechains as determined by a structure search in each sidechain for a basic set of small fragments. Lastly, in Fig \ref{fig:figure2} (D), the model learns to generate unnatural proteins of similar atom number and backbone size to the training proteins-- but does tend to slightly underestimate the size of both.

\begin{figure*}[t]
    \centering
    \includegraphics[width=0.99\textwidth]{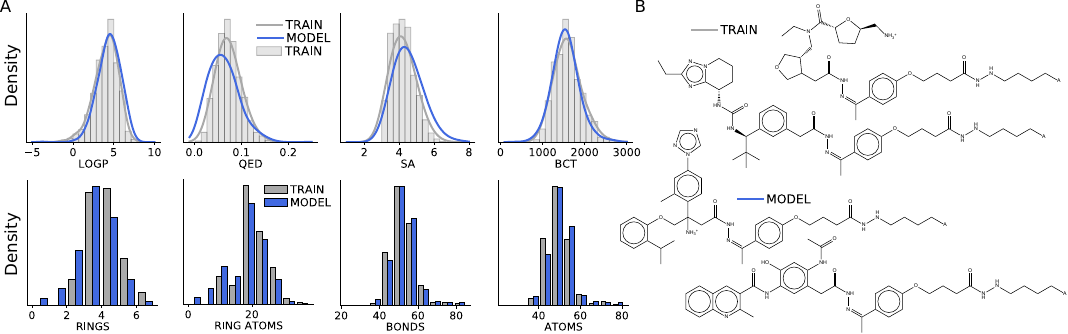}
    \caption{\textbf{Antibody Drug Conjugates-- Warheads}  
    (A) Histograms and density plots of atom-level measures or properties of warheads. 
    Density plots use a Gaussian kernel density estimator fit to LogP, MW, and PSA values for the training molecules by tuning the bandwidth parameter.
    (B) Examples of model and train "warheads".
    }
    \label{fig:figure4}
\end{figure*}

\subsection{Antibody Drug Conjugates}

Next, we test the ability of the language model to generate proteins attached to small molecules and simultaneously explore protein space and chemical space. One of the most promising examples of this kind of biomolecule with immense therapeutic potential are antibody-drug conjugates, which are a form of cancer therapy intended to target and kill cancer cells but spare healthy cells \cite{beck2017strategies}. Structurally, they are composed of an antibody attached to single or multiple anticancer drugs typically using some linker molecule. To construct a synthetic dataset of antibody-drug conjugates, as shown in Fig \ref{fig:figure3} (A), we attach a single drug-like molecule from the ZINC dataset \cite{irwin2005zinc} to single-domain antibodies (sdAbs) from the structural antibody dataset \cite{schneider2022sabdab, constantinnano} in order to test the ability of language models to generate antibody-drug conjugates. We use two possible linkers for cysteine attachments and two other linkers for lysine attachments. Both linkers are selected from real antibody-drug conjugates described in \cite{beck2017strategies}. The linker is randomly attached to the small molecule from ZINC and the specific lysine or cysteine residue for attachment is also randomly chosen. Since there are only ~1K sdAbs in the structural antibody dataset we use data augmentation to expand the dataset size to ~250K proteins that can be attached to every molecule in ZINC. 

After training, we again generate 1K samples from the language model for evaluation, We first test the model's ability to explore protein space and learn the distribution of single-domain antibodies-- the results are shown in Fig \ref{fig:figure3} (B-D). Similar to the standard protein data, we compare the distribution of amino acids in the training sequences to the distribution learned by the model. We plot histograms, in Fig. \ref{fig:figure4}(B), displaying the frequency of occurrence of every amino acid in samples from both the model and the training data-- from these, we can see the language model accurately learns the training distribution of amino acids. Similarly, in Fig. \ref{fig:figure4}(C), the model accurately learns the size of the training sdAbs. 

Similar to the standard proteins, we can attempt to determine the amino acid sequences of the single-domain antibodies (ignoring the warheads). We determine roughly $\sim$ 90.8\% of samples are proteins and their primary sequences are entirely unique and novel (there are no duplicates and all are different from training sequences).

Even further, Alphafold structure predictions \cite{jumper2021highly}, visualized in Fig. \ref{fig:figure3}(D) and Supplementary Fig. \ref{fig:sfigure7}, confidently show that the language model produces sequences that fold into the expected structure for single domain antibodies.
Additionally, based on the primary sequence overlap of model samples with their nearest PDB training example in Fig. \ref{fig:figure1}(C) and Supplementary Fig. \ref{fig:sfigure1}, the model, without memorizing, learns the amino acid sequence structure that defines the training sdAbs. The primary sequence overlap with training examples ranges from 63\% to 93\% in the supplementary information. Investigating further, we see that the model draws heavily from the sdAb sequences making new examples of sequences by memorizing small snippets of amino acids and using larger training snippets but with a large number of single mutations randomly distributed throughout the snippet. 
 
From the training examples and model samples, we detach and collect "warheads" which we expand the definition of to include the linker and sidechain in addition to the small molecule (warhead typically refers to just the small molecule). In Fig. \ref{fig:figure4} (B), two examples of train and model warheads are shown as graphs to clarify this. Additional model and training warheads are shown as graphs in Supplementary Fig. \ref{fig:sfigure8model} and \ref{fig:sfigure8train}-- as expected the same four linkers repeat across samples but the small molecules attached to them differ and are structurally similar to the ZINC molecules in the training warheads. 

We also evaluate the language model's warheads in terms of their atom-level properties.
In Fig \ref{fig:figure4} (A), the model captures the atom-level properties of the training warheads, specifically, it learns the continuous atom-level properties of the training warheads including LogP \cite{wildman1999prediction}, drug-likeness (QED) \cite{bickerton2012quantifying}, Synthetic Accessibility Score (SA) and molecular graph complexity (BCT) as well as the number of atoms, bonds, rings and atoms in rings. However, the model slightly underestimates the main modes for QED and SA as well as the number of rings per warhead.

Additionally, we assess the model warheads and compare them with the training warheads, we find that model warheads are unique (there are no duplicates and the model is not repeating a few examples) as well as novel (the model does not make exact copies of warheads from the training data). Given that the linkers are memorized, this indicates that the model is learning to generate new small molecules similar to ZINC molecules and effectively exploring chemical space at the same time it learns to explore the protein space defined by the sdAbs.

Also, in Supplementary Fig. \ref{fig:sfigure2}, we see that the model does learn the atom-level properties of the training antibody drug-conjugates. Additionally, in Supplementary Fig. \ref{fig:mndc1}-\ref{fig:mndc3}, we show a single train antibody drug-conjugate and four model samples. 

\section*{Discussion}

In this work, we show that chemical language models can generate, atom by atom, entire proteins, unnatural proteins, and protein drug conjugates. By analyzing generated samples we find that language models learn multiple hierarchical layers of molecular information that define the training biomolecules. This includes atom-level molecular properties or residue-level constraints for backbone and amino acid structure as well as primary sequence patterns and motifs that define meaningful secondary and tertiary structure. Indeed, chemical language models learn to generate protein structures as sequence representations of atom-level graphs that are similar to the training proteins in the PDB.
 
Effectively we demonstrate that chemical language models can also serve as biological language models-- capable of learning the language of proteins atom by atom. Importantly, in contrast to protein language models that only learn representations of amino acid sequences, chemical language models generate entire molecular graphs, and because of this we are able to show that language models can be used to explore not just chemical space and in between chemical and protein space but also protein space itself, beyond protein space, or even both chemical and protein space at the same time. 

Further work should be done to ensure the model generates valid protein structure including correct backbone and amino acid form. 
This will also assist the model in learning distributions consisting of larger biomolecules including structures with more than 150 residues and multiple domains.
Using memorizing Transformers \cite{wu2021memorizing} may help the model generate valid protein sequences. Also, other architectures built for longer sequence lengths \cite{child2019generating} can increase the size and range of structures that the model can learn. Another limitation is that we do not  consider the three-dimensional structure of the biomolecules and generate atom-level sequence representations. This problem can not be easily rectified because no training data with 3D information exists for unnatural proteins and protein-drug conjugates. A potential solution that does not requires training data would be to use reinforcement learning 
\cite{flam2022scalable} or bayesian optimization \cite{wu2020bayesian} and guide the model to generate 3D structure using energy.

The goal of this work is to demonstrate the power of chemical language models and their ability to learn atom-level representations of biomolecules. We envision future language models will be able to explore any combinatorial space in chemistry or biology using any representation type the user wishes \cite{flam2023language}.




\section{Methods}

\subsection{Datasets}
From the PDB we successfully parse around $\sim$10K proteins between 50 and 150 residues. In all datasets, we only parse proteins that conform to atom-level graphs with no more than 2 macrocycles (created by residue-residue connections) -- this makes primary sequence determination more successful.  Given this constraint, we parse around $\sim$10K and $\sim$5K proteins from the PDB for the first two training datasets. In order to increase the size of the training data, we randomize the atom orderings of each protein in \texttt{rdkit}  to obtain multiple different random copies of each biomolecule as SMILES (and then SELFIES strings). Using this data augmentation we expand all training datasets to around $\sim$250K sequences. We use \texttt{rdkit} \cite{landrum2013rdkit} to represent each protein as atom-level graphs and make side-chain modifications.
We also made use of Colabfold \cite{mirdita2022colabfold} for quick visualization and NGLview \cite{nguyen2018nglview} for figure construction. 

\subsection{Tokenization}
We use SELFIES \cite{krenn2019selfies} version one for the sequence representation of atom-level protein graphs. 
Other than special tokens like  $ \texttt{[BOS],[EOS],[PAD],[UNK]} $, the vocab $\mathcal{T}$ consists of standard selfies tokens, encoding all information in a molecular graph including: 
atom tokens $\{ \texttt{[C]},\texttt{[N]}, \dots \}$, 
bond tokens $\{ \texttt{[=C]},\texttt{[\#N]}, \dots \} $, 
ring tokens: $\{ \texttt{[Ring1]},\texttt{[Ring2]}, \dots \} $  
branching tokens: $\{ \texttt{[Branch1\_1]},\texttt{[Branch1\_2]}, \dots \} $. 
In total for all datasets, the vocabulary is around $\sim$30 tokens.

\subsection{Language Modeling for Molecular Design} 
In language modeling for molecular design, we want to estimate the unsupervised distribution of the training molecules $(\textsc{mol}_1, \textsc{mol}_2,\dots , \textsc{mol}_n)$ each composed of variable length sequences of tokens from a chemical language $[\texttt{CT}]_i$ where $\texttt{CT}\in \mathcal{T}$ such that $\textsc{mol}=([\texttt{CT}]_1, [\texttt{CT}]_2,\dots , [\texttt{CT}]_n)$. The joint probabilities over a single molecule can be written as
\begin{align}\label{eq:mol}
    p(\textsc{mol})=\prod _{i=1}^n p([\texttt{CT}]_n | [\texttt{CT}]_{n-1},\dots [\texttt{CT}]_1)
\end{align}
These probabilities $p([\texttt{CT}]_n \ | \ [\texttt{CT}]_{n-1},\dots [\texttt{CT}]_1)$ are modeled using a Transformer \cite{radford2018improving} that is trained using stochastic gradient descent. 

\subsection{Training}
During training, we one-hot encode SELFIES sequences using a basic vocabulary that consists of ~30 possible alphabet tokens. All language models are trained using next-token prediction conditioned on the entire sequence for context. The training data only uses sequences that have a maximum length of 1664 tokens. 
We trained language models with decoder only, GPT-like architecture \cite{radford2019language} with 4 attention heads and between 1 and 10 Million parameters. Language models are implemented in Python 3 with PyTorch \cite{paszke2019pytorch}. Molecules properties are computed using  \texttt{rdkit}  \cite{landrum2013rdkit}.

\section{Acknowledgements}   A.A.-G. acknowledge funding from Dr. Anders G. Fr{\o}seth.
A.A.-G. also acknowledges support from the Canada 150 Research Chairs Program, the Canada Industrial Research Chair Program, and from Google, Inc.
Models were trained using the Canada Computing Systems \cite{baldwin2012compute}. This research was undertaken thanks in part to funding provided to the University of Toronto's Acceleration Consortium from the Canada First Research Excellence Fund.

\bibliographystyle{apsrev4-1}
\bibliography{main}

\pagebreak
\begin{widetext}

\newpage

\section*{Supplementary Information}

\setcounter{equation}{0}
\setcounter{figure}{0}
\setcounter{table}{0}
\makeatletter
\renewcommand{\theequation}{S\arabic{equation}}
\renewcommand{\thefigure}{S\arabic{figure}}

\begin{figure*}[h]
    \centering
    \includegraphics[width=0.9\textwidth]{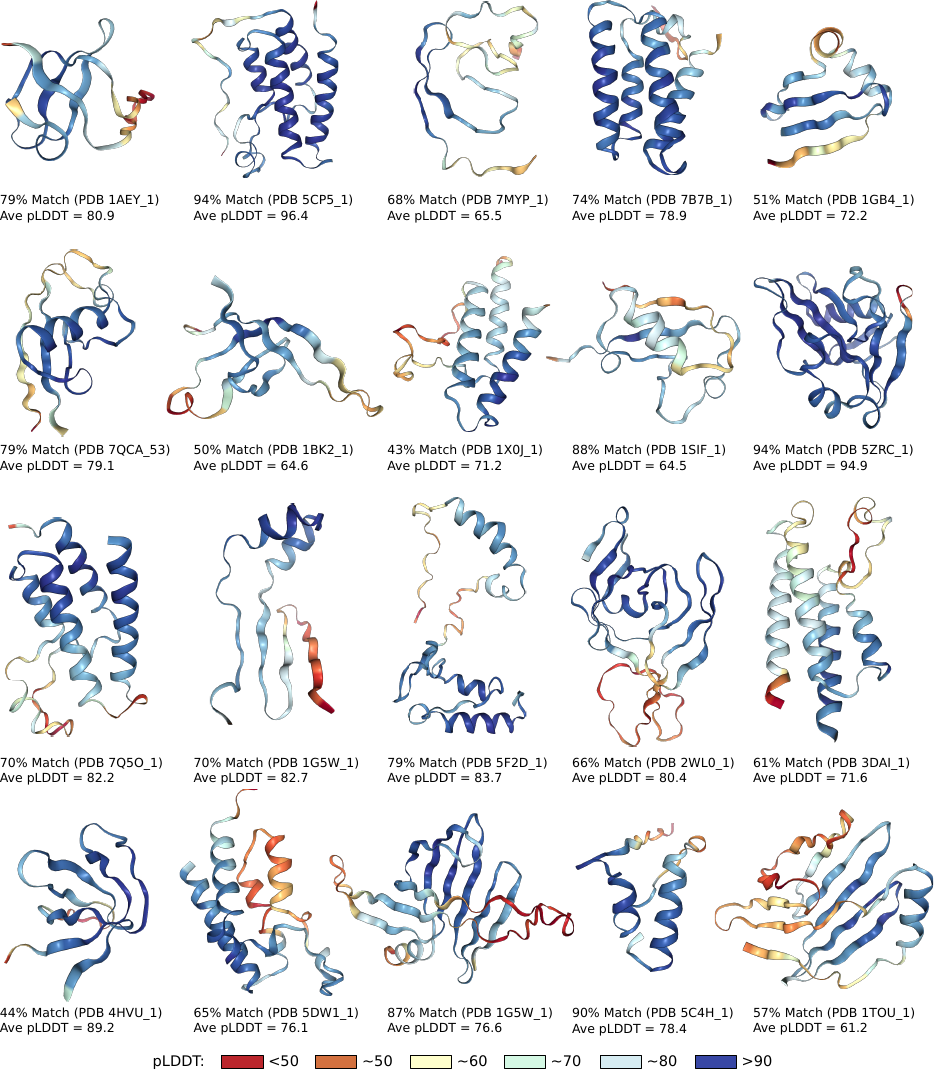}
    \caption{Examples of proteins generated by the model visualized by Alphafold and colored by pLDDT.}
    \label{fig:sfigure1}
\end{figure*}

\begin{figure*}[t]
    \centering
    \includegraphics[width=0.99\textwidth]{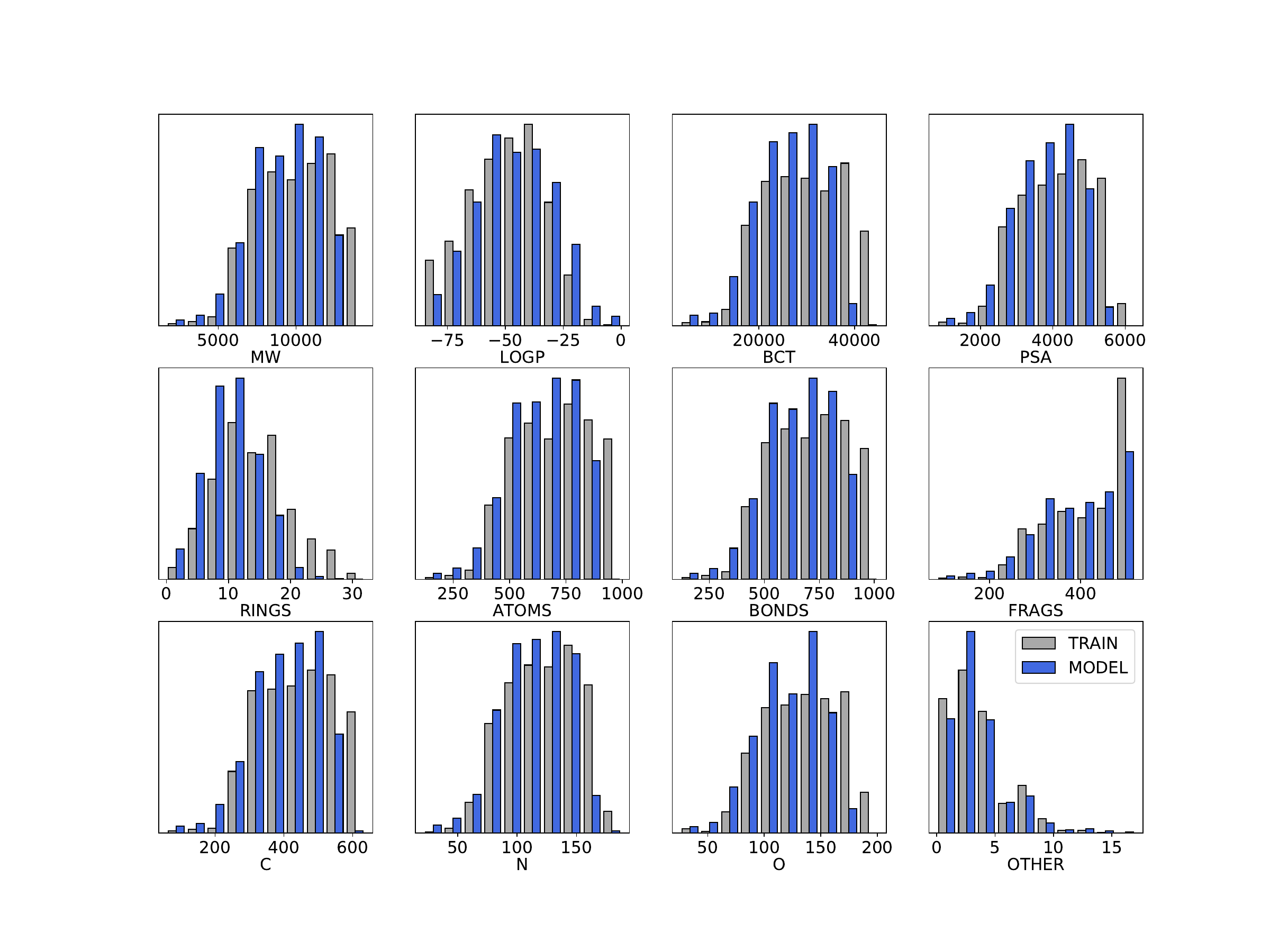}
    \caption{Histograms of atom level properties for the basic proteins training data. These include exact molecular weight (MW),
     octanol-water partition coefficient (LogP) \cite{wildman1999prediction}, molecular complexity (BCT), topological polar surface area (PSA), number of rings (RINGS), number of atoms (ATOMS), number of bonds (BONDS), number of fragments found by breaking up the molecule at rotatable bonds (FRAGS), number of carbons (C), number of nitrogens (N), number of oxygens (O), and number of any other atoms (OTHER).   
     }
    \label{fig:sfigure2}
\end{figure*}

\begin{figure*}[t]
    \centering
    \includegraphics[width=0.7\textwidth]{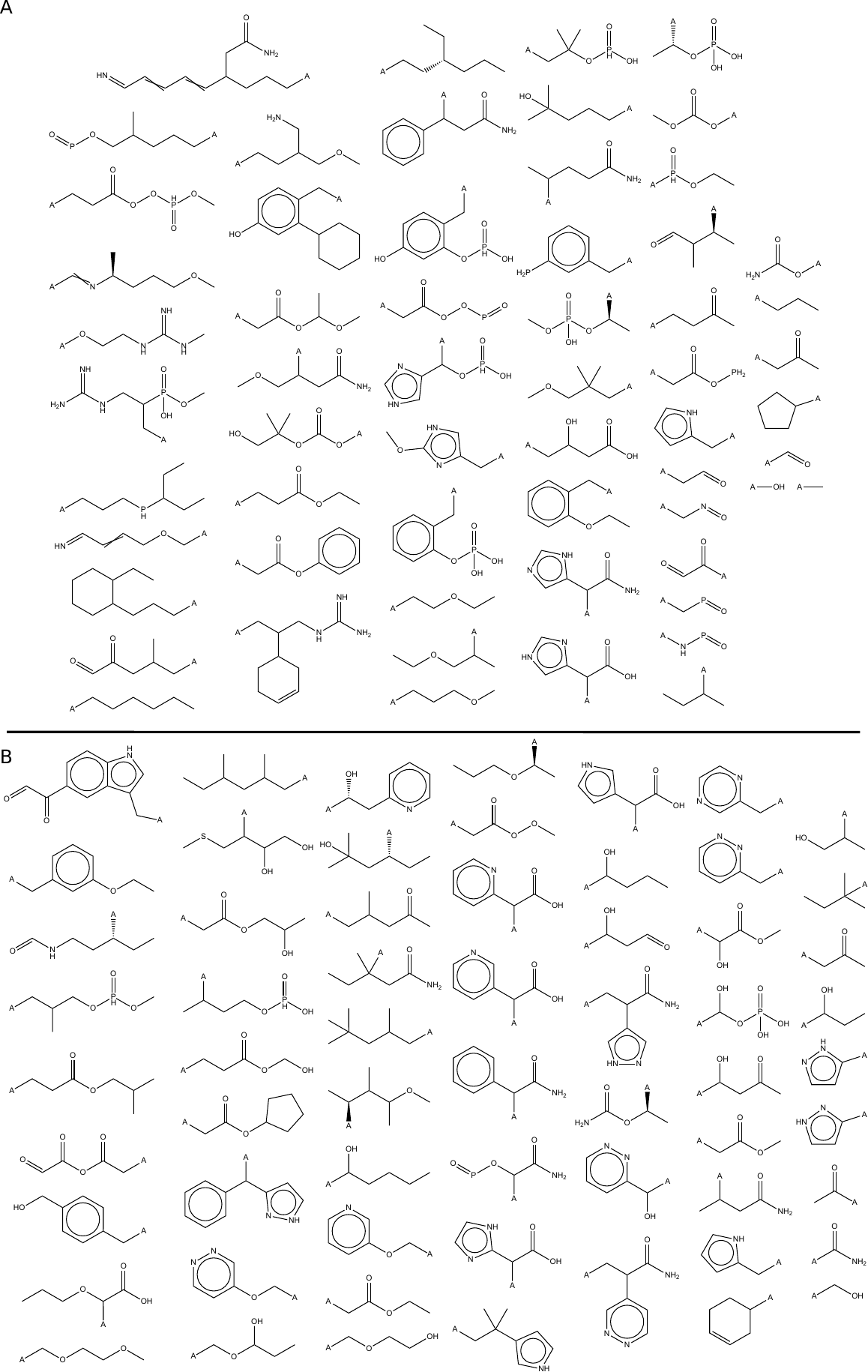}
    \caption{Examples of side chains from a single training protein with unnatural amino acids and a single model generated protein with unnatural amino acids.}
    \label{fig:sfigure3}
\end{figure*}

\begin{figure*}[t]
    \centering
    \includegraphics[width=0.9\textwidth]{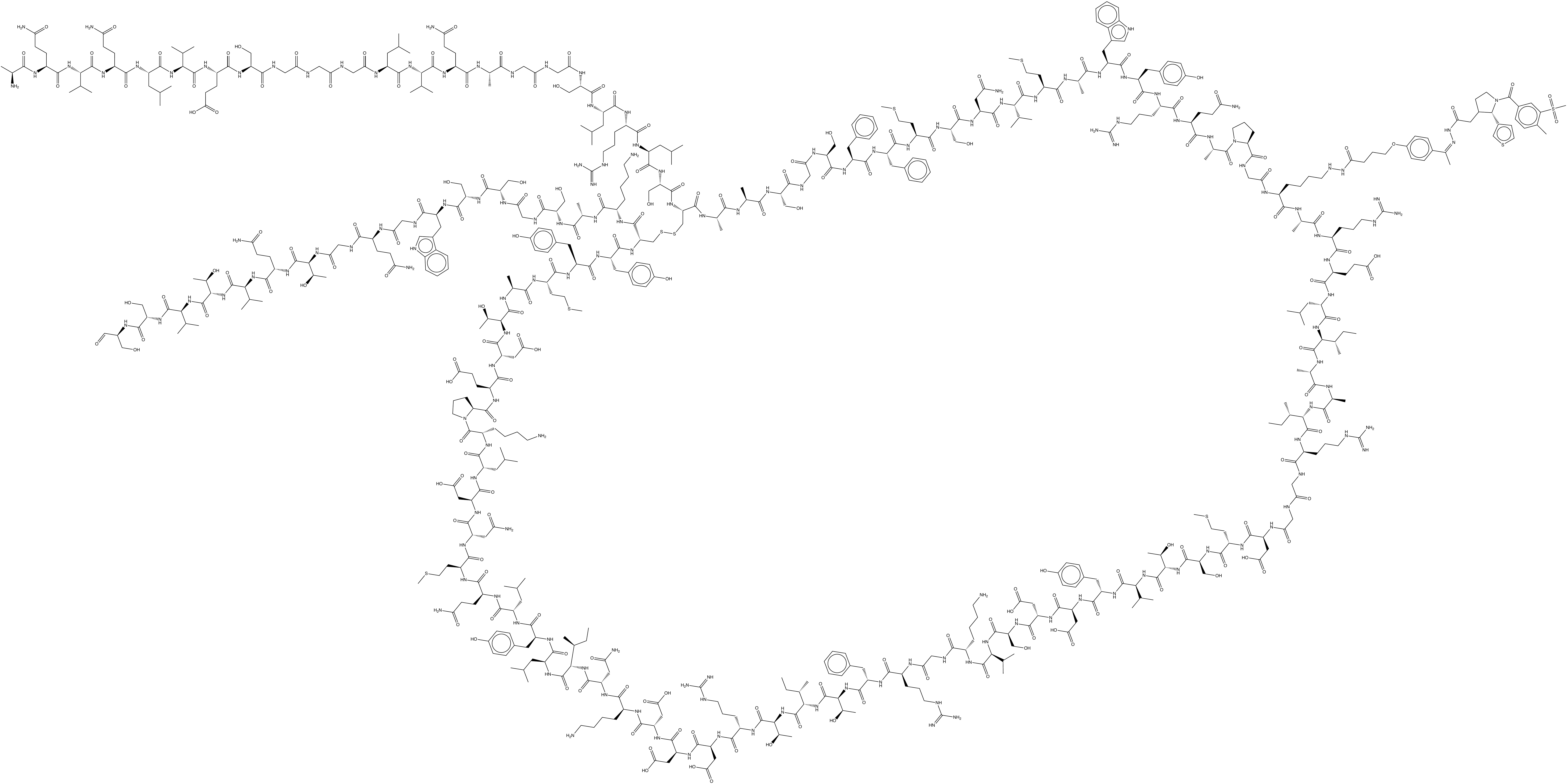}
    \includegraphics[width=0.9\textwidth]{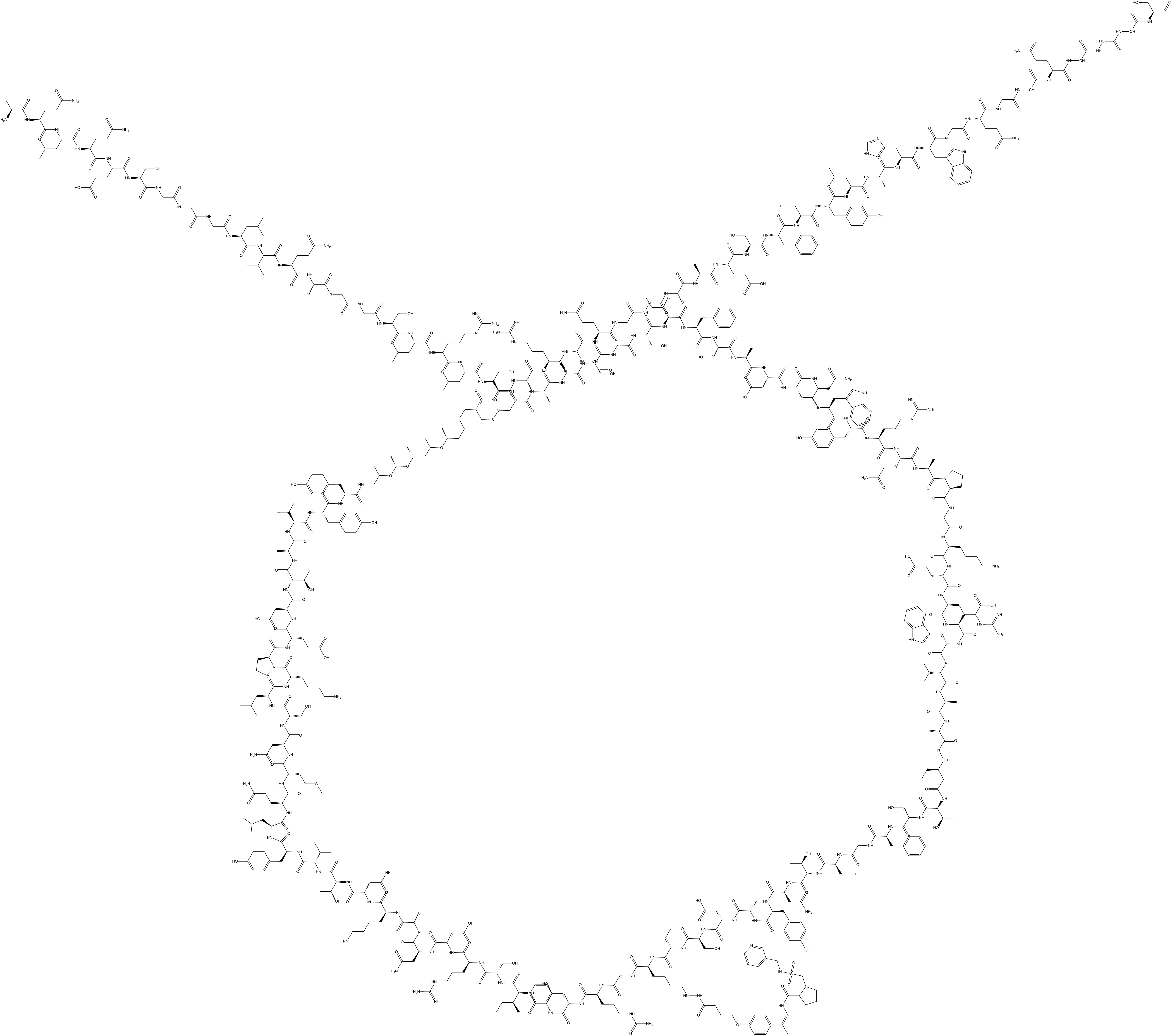}
    \caption{\textbf{Antibody-Drug conjugates} Example antibody-drug conjugates from the training data (above) and one example produced by the language model (below) and plotted using ChemDraw as an atom-level graph. 
    }
    \label{fig:mndc1}
\end{figure*}

\begin{figure*}[t]
    \centering
    \includegraphics[width=0.99\textwidth]{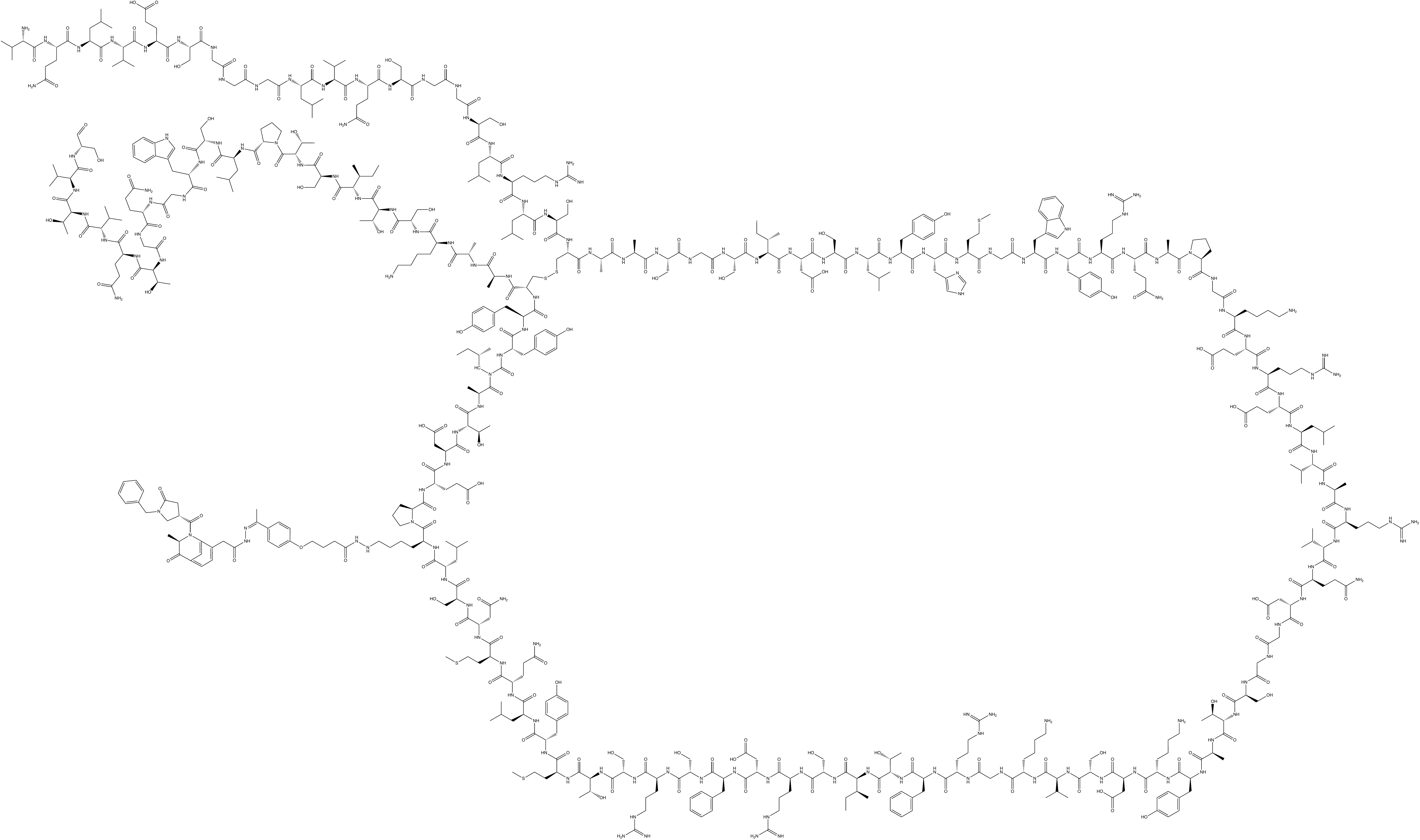}
    \includegraphics[width=0.99\textwidth]{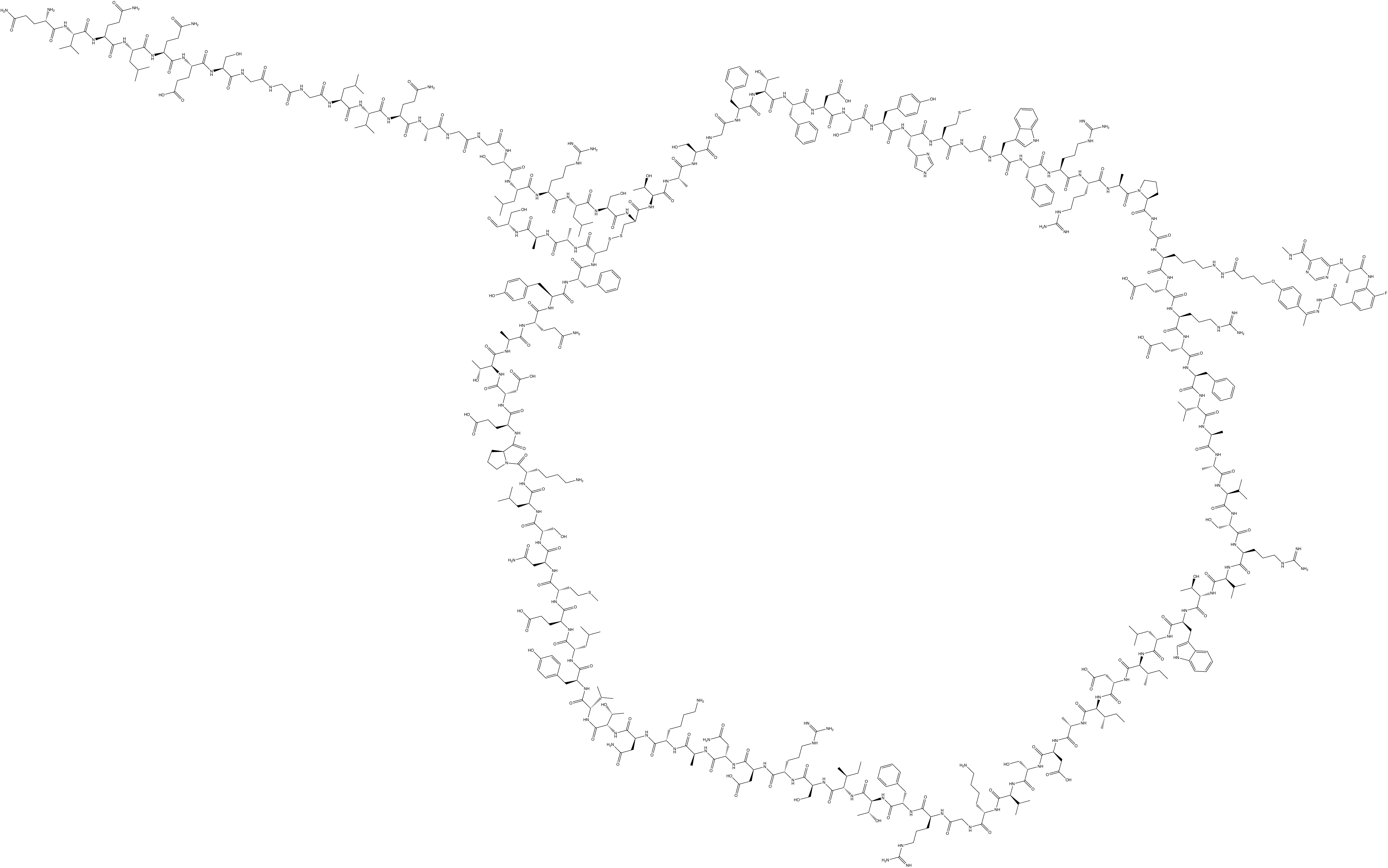}
    \caption{\textbf{Model Antibody-drug conjugates} Example antibody-drug conjugates produced by the language model and plotted using ChemDraw as an atom-level graph. 
    }
    \label{fig:mndc2}
\end{figure*}

\begin{figure*}[t]
    \centering
    \includegraphics[width=0.9\textwidth]{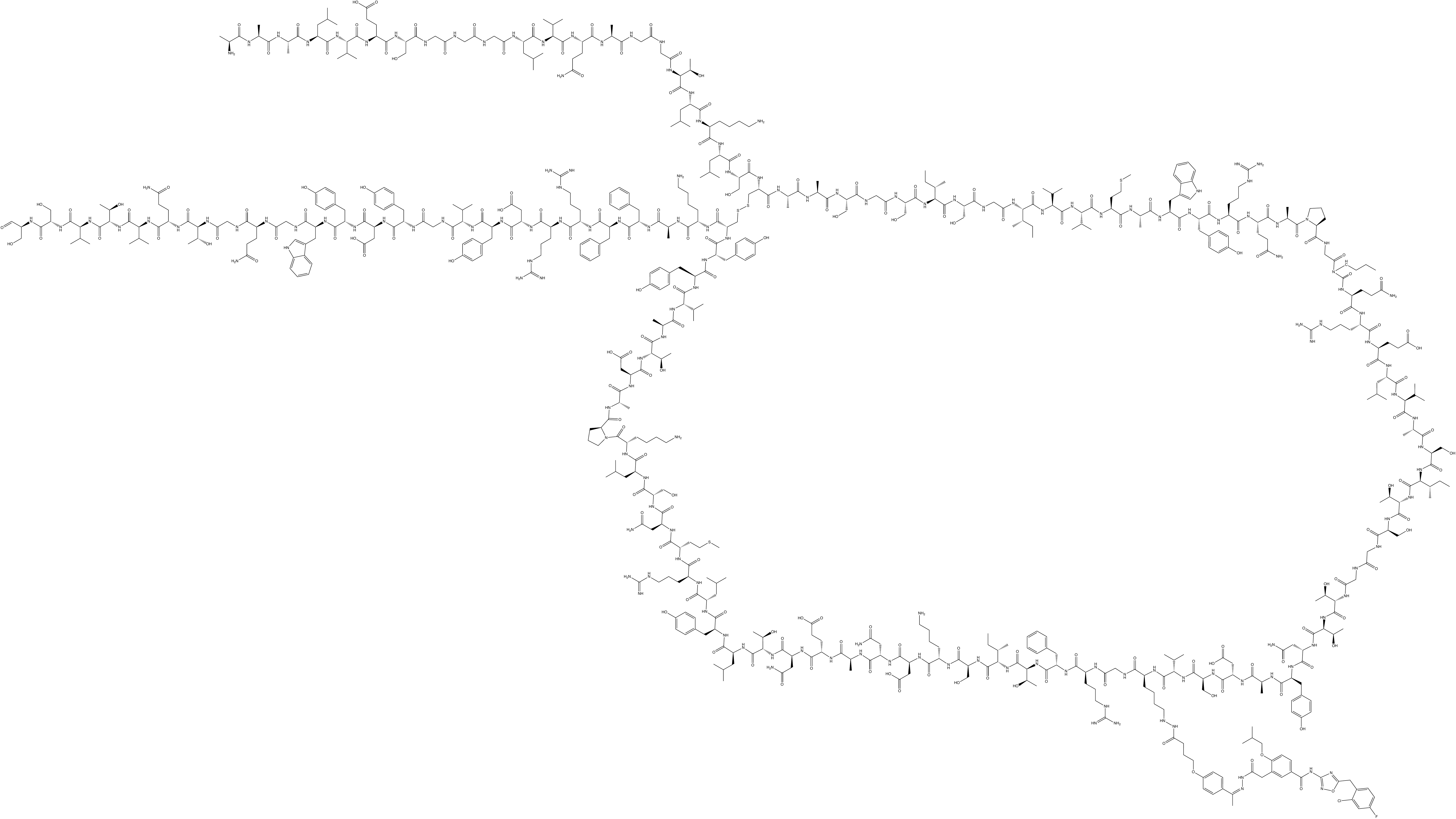}
    \includegraphics[width=0.9\textwidth]{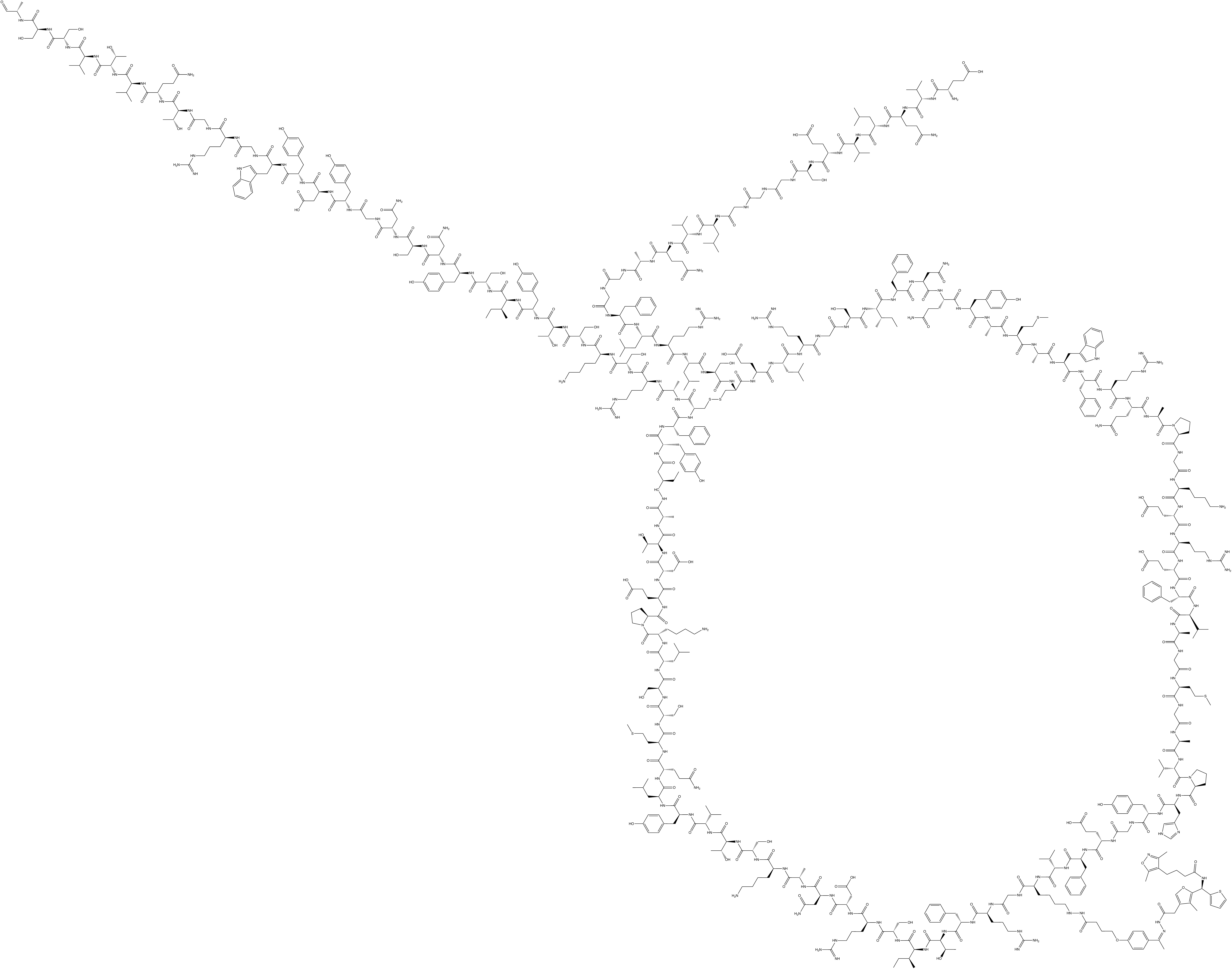}
    \caption{\textbf{Model Antibody-drug conjugates} Example antibody-drug conjugates produced by the language model and plotted using ChemDraw as an atom-level graph. 
    }
    \label{fig:mndc3}
\end{figure*}

\begin{figure*}[t]
    \centering
    \includegraphics[width=0.99\textwidth]{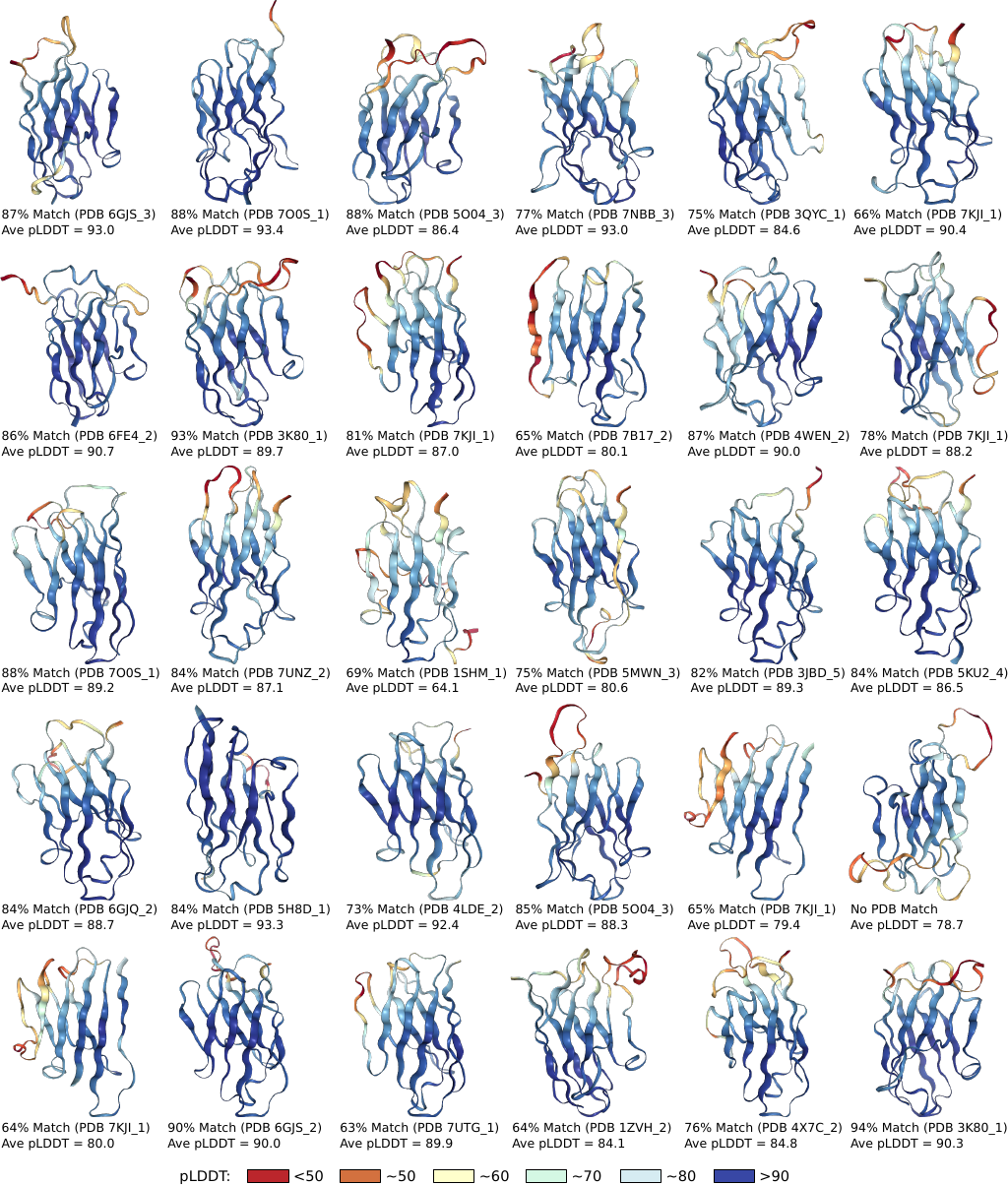}
    \caption{Examples of single domain antibodies from samples of sdAb-drug conjugates (excluding warheads) generated by the model visualized by Alphafold and colored by pLDDT.}
    \label{fig:sfigure7}
\end{figure*}

\begin{figure*}[t]
    \centering
    \includegraphics[width=0.85\textwidth]{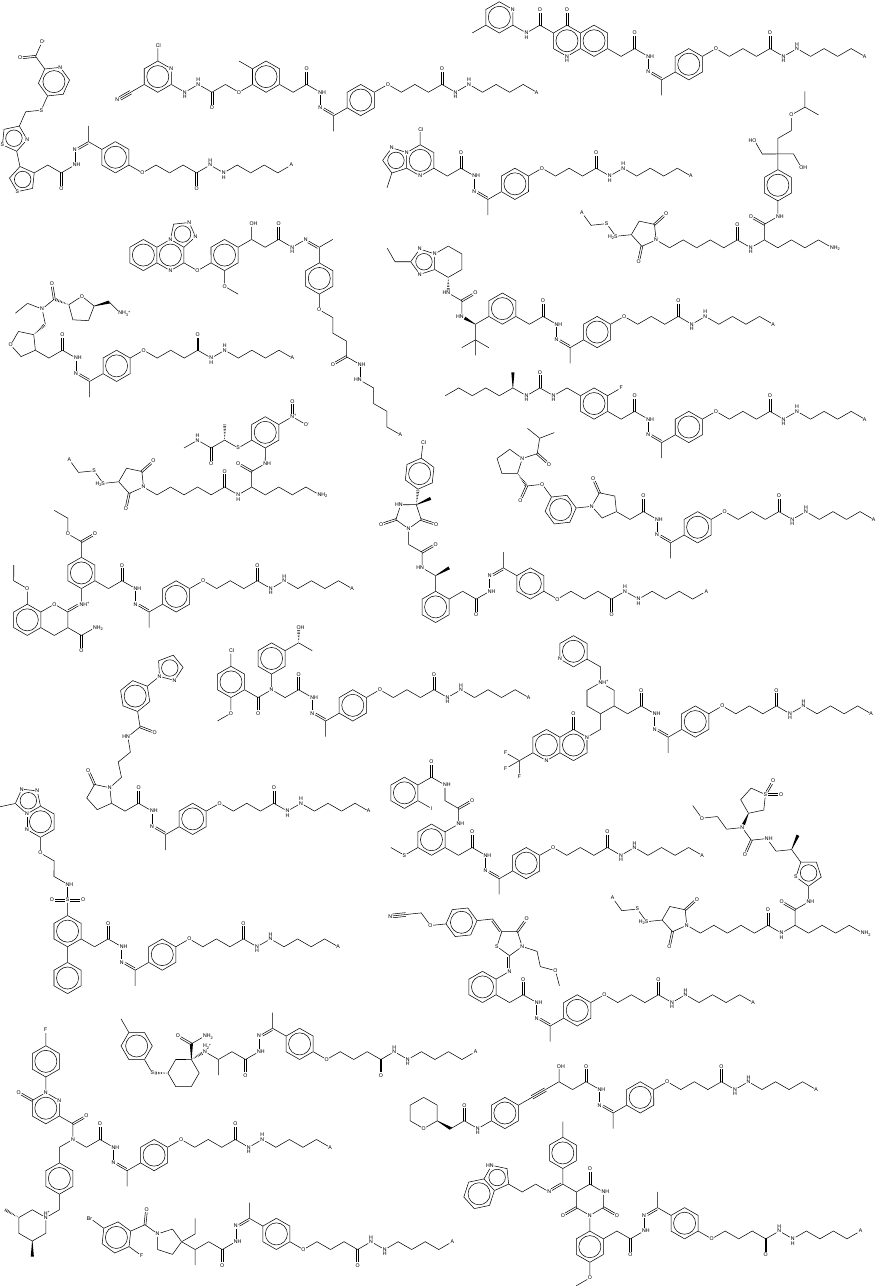}
    \caption{Examples of detached warheads from training single domain antibody-drug conjugates.}
    \label{fig:sfigure8train}
\end{figure*}

\begin{figure*}[t]
    \centering
    \includegraphics[width=0.85\textwidth]{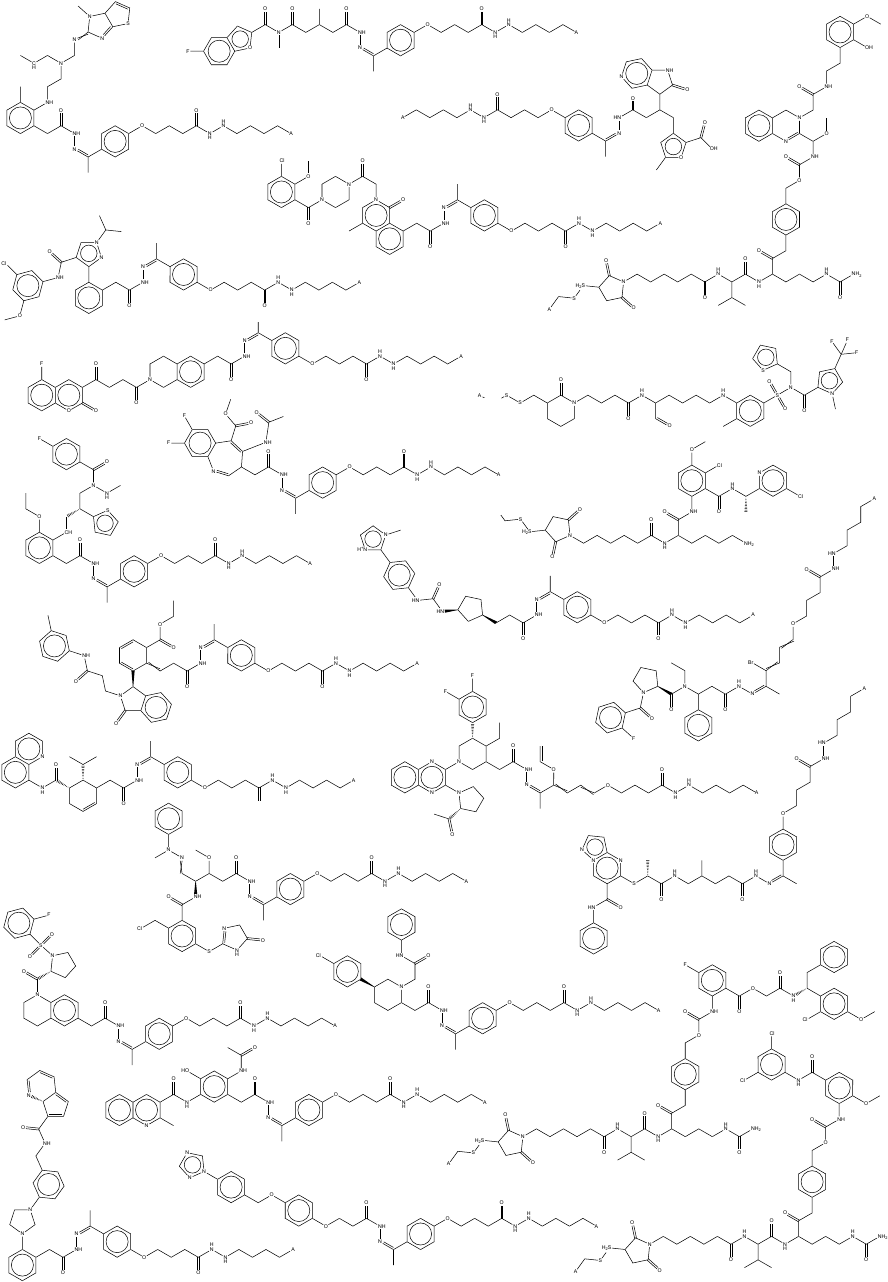}
    \caption{Examples of detached warheads from model generated single domain antibody-drug conjugates.}
    \label{fig:sfigure8model}
\end{figure*}

\begin{figure*}[t]
    \centering
    \includegraphics[width=0.99\textwidth]{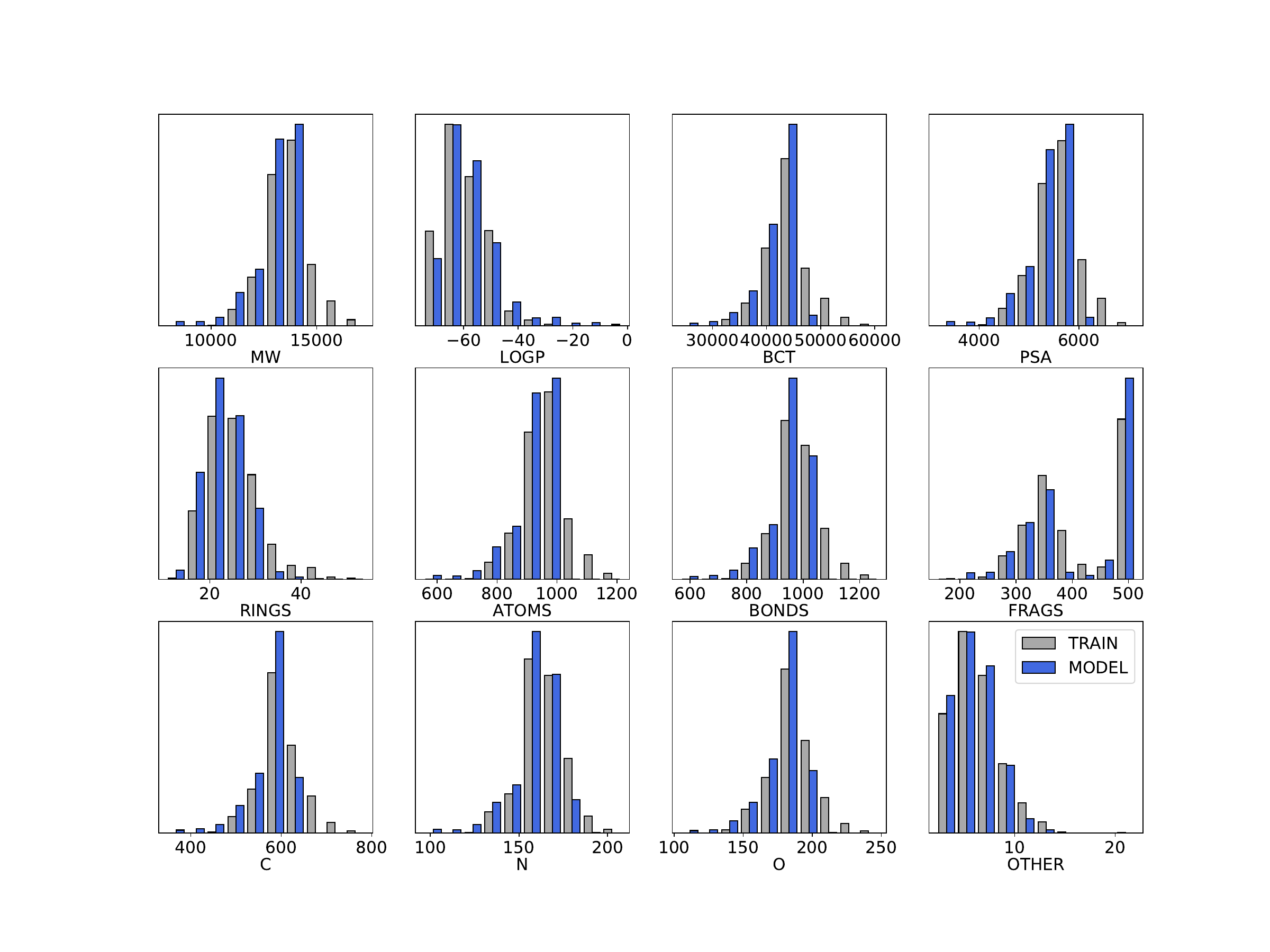}
    \caption{Histograms of atom level properties for the antibody-drug-conjugates training data. These include exact molecular weight (MW),
     octanol-water partition coefficient (LogP) \cite{wildman1999prediction}, molecular complexity (BCT), topological polar surface area (PSA), number of rings (RINGS), number of atoms (ATOMS), number of bonds (BONDS), number of fragments found by breaking up the molecule at rotatable bonds (FRAGS), number of carbons (C), number of nitrogens (N), number of oxygens (O), and number of any other atoms (OTHER)}
    \label{fig:sfigure9}
\end{figure*}

\end{widetext}

\end{document}